\documentclass[aps, prd,preprint,showpacs,superscriptaddress,groupedaddress,nofootinbib]{revtex4-1}
\usepackage{amsmath,amsfonts}
\usepackage[dvips]{graphicx}
\usepackage{morefloats}
\graphicspath{{Fig/}}
\begin{document}

\title{On Radiative Fluids in Anisotropic Spacetimes}
\author{Dmitry Shogin}
\email{dmitry.shogin@uis.no}
\affiliation{Faculty of Science and Technology, University of Stavanger, N-4036 Stavanger, Norway}
\author{Per Amund Amundsen}
\email{per.a.amundsen@uis.no}
\affiliation{Faculty of Science and Technology, University of Stavanger, N-4036 Stavanger, Norway}

\begin{abstract}
We apply the second-order Israel-Stewart theory of relativistic fluid- and thermodynamics to a physically realistic model of a radiative fluid in a simple anisotropic cosmological background. We investigate the asymptotic future of the resulting cosmological model and review the role of the dissipative phenomena in the early Universe. We demonstrate that the transport properties of the fluid alone, if described appropriately, do not explain the presently observed accelerated expansion of the Universe. Also, we show that, in constrast to the mathematical fluid models widely used before, the radiative fluid does approach local thermal equilibrium at late times, although very slowly, due to the cosmological expansion.
\end{abstract}


\maketitle

%%%%%%%%%%%%%%%%%%%%%%%%%%%%%%%%%%%%%%%%%%%%%%%%%%%%%%%%%

\section*{INTRODUCTION}
Cosmological fluids are commonly assumed to be perfect. This implies that the fluids are in the state of local thermal equilibrium, manifest no dissipative effects and, therefore, do not generate entropy. However, although the perfect fluid assumption leads to mathematical simplificity and has been successfully used in many situations, its area of application is restricted. As follows from the kinetic theory, all real substances \textit{do} have transport properties such as viscosity and thermal conductivity; this leads to dissipation, and, therefore, to \textit {irreversible} fluid dynamics. Moreover, a fluid without transport properties would never approach the local equilibrium state: the forces, which drive the fluid towards equilibrium, are always dissipative. Dissipative effects must have played a crucial role in the early Universe, when the processes of inflation and reheating, involving interactions between cosmological fluids of different kinds, took place.
\par 
A brief review of the traditional theories of dissipative relativistic fluid- and thermodynamics can be found in~\cite{Maartens1996}; see references therein for applications to cosmology. The most advanced approach to date is the second-order Israel-Stewart (IS) theory \cite{Israel1976,Israel1979}, often referred to as transient, or causal, thermodynamics. It should be noted that the full version of the IS theory must be preferred over its widely used truncated version; the latter produces substantially distinct results and fails in multiple aspects, see e.\,g. \cite{Shogin2015,Shogin2016}.
\par 
The standard spatially flat, homogeneous and isotropic Friedmann-Robertson-Walker cosmological models have been studied using the full IS theory~\cite{Maartens1995,Coley1996}. In particular, Maartens~\cite{Maartens1995} discussed bulk viscous inflation. However, the simple geometry of the standard model does not allow for dissipative mechanisms other than bulk viscosity: anisotropic backgrounds must be considered when modelling realistic dissipation.
\par 
The applications of causal thermodynamics to anisotropic cosmological models started with the pioneering work by Belinskii et al.\,~\cite{Belinskii1979}. Later, van den Hoogen and Coley \cite{Hoogen1995} considered Bianchi type~V cosmological models using the truncated version of the IS theory. Recently, Shogin et al. studied Bianchi type~I~\cite{Shogin2016}, IV and V~\cite{Shogin2015} cosmological models. Inter alia, it has been demonstrated that the IS theory can break down in cosmological applications, as viscous stresses drive the fluid essentially far from local thermal equilibrium.
\par 
For the sake of convenience, simplified \textit{mathematical} fluid models with phenomenological expressions for the transport coefficients have been used. In the present research we make a step further and consider a \textit{physics-based} model for a two-component radiative fluid in Bianchi type~I spatially anisotropic cosmological backgrounds.
\par 
Finally, we make a remark about the notations we use throughout the paper. Small Greek indices refer to the four-dimensional spacetime and run from~0 to~3. Small Latin indices refer to the three-dimensional spatial section and run from~1 to~3. Also, we accept the Einstein summation convention and the notation of the covariant derivative by semicolons. Various physical and geometrical quantities are introduced in the text of the paper.
%%%%%%%%%%%%%%%%%%%%%%%%%%%%%%%%%%%%%%%%%%%%%%%%%%%%%%%%

\section*{THE EINSTEIN FIELD EQUATIONS}
The Einstein field equations are the fundamental equations of general relativity, describing the interaction between the geometry and the energy-matter content of the spacetime. In tensor form, they are written as:
\begin{equation}
\label{Eq:EFE:EFE}
R_{\alpha \beta}-\frac{1}{2}Rg_{\alpha \beta}=T_{\alpha \beta}.
\end{equation} 
The left-hand side of~(\ref{Eq:EFE:EFE}) expresses the spacetime geometry in terms of the metric tensor~$g_{\alpha \beta}$, the Riemann curvature tensor~$R_{\alpha \beta}$ and the Ricci scalar~$R$. The matter content of the cosmological model enters the equations as the energy-momentum tensor~$T_{\alpha \beta}$ on the right-hand side of~(\ref{Eq:EFE:EFE}).
\par 
The field equations are subject to geometrical restrictions, known as the Bianchi identities:
\begin{equation}
\label{Eq:EFE:Bianchi}
T^{\alpha \beta}_{\phantom{\alpha}\phantom{\beta};\beta}=0.
\end{equation}
Physically, the Bianchi identities represent the energy and momentum conservation laws.
\par 
The orthonormal frame approach~\cite{Elst1997} allows to rewrite the tensor equations~(\ref{Eq:EFE:EFE})-(\ref{Eq:EFE:Bianchi}) as equations in terms of scalar quantities using the commutation functions as variables, see e.\,g.~\cite{Wainwright1997} for further details. The Hubble scalar~$H$ and the geometric rate of shear tensor~$\sigma_{ab}$, arising in this procedure, are important kinematic quantities. Their physical meaning becomes clear if one chooses a spherical space element in the comoving frame. Under the action of~$H$, the volume of the element changes in time, but not the spherical shape. Under the action of~$\sigma_{ab}$, the spherical shape of the element is being distorted, the volume being unchanged.

%%%%%%%%%%%%%%%%%%%%%%%%%%%%%%%%%%%%%%%%%%%%%%%%%%%%%%%%%

\section*{THE PHYSICAL FLUID MODEL}

In the present research, we consider specific two-component fluids which are of particular interest for cosmological applications~\cite{Schweizer1982}. The first component is some material medium with short mean free paths, which is assumed to be locally in thermal equilibrium. The second component, consisting of radiation quanta (photons) with finite mean free paths, is in an off-equilibrium state, but close to equilibrium with the material medium.
\par 
We use the geometrized units throughout the paper, i.\,e. the reduced Planck's constant, the speed of light in vacuum, the Boltzmann's constant and the Einstein's gravitational constant are chosen to equal unity:
\begin{equation}
\hbar=c=k_B=4\pi G=1.
\end{equation}
With this choice, the dimensions of all the variables become integer power of length.
\par 
The fluid is assumed to obey the following equations of state:
\begin{align}
\rho(n,T) &= mn+\frac{1}{3}nT+aT^4,\\
p(n,T) &= nT+\frac{1}{3}aT^4,
\end{align} 
where~$m$ stands for the (constant) mass of the material particles, $n$ for their number density, and $a=\pi^2/15$ is the radiation constant for photons. The temperature variable~$T$ represents the {\it actually observed} temperature of the fluid mixture (the so-called Eckart temperature), which is, in general, different from the equilibrum temperature of the material medium.
\par 
The fluid mixture behaves like a relativistic imperfect fluid, the energy-momentum tensor~$T_{\alpha \beta}$ of which can be decomposed as follows:
\begin{equation}
T_{\alpha \beta}=(\rho+p+\pi) u_\alpha u_\beta + (p+\pi)g_{\alpha \beta} +q_{(\alpha} u_{\beta)}+\tau_{\alpha \beta}.
\end{equation}

The quantities $\pi, q_\alpha$ and $\tau_{\alpha \beta}$ represent the scalar, vector, and tensor dissipative fluxes, respectively:~$\pi$ can be identified as the bulk viscosity; $q^\alpha$, with $q_\alpha u^\alpha =0$, as the heat flux vector; and $\tau_{\alpha \beta}$, with $\tau_{\alpha \beta} u^\beta = \tau_\alpha^{~\alpha}=0$, as the anisotropic stress tensor (shear viscosity). We choose the fluid four-velocity vector $u_\alpha$ to be the particle four-velocity of the material medium (this choice is known as the Eckart frame), and assume the two-component fluid to be non-tilted, so that $u^\alpha$ is orthogonal to the spatial hypersurface. Furthermore, we neglect the effects of thermal conductivity, which, together with the previous assumption, results in $q^\alpha \equiv 0$. This leads to significant mathematical simplifications: in particular, the geometric rate of shear tensor and the anisotropic stress tensor take the diagonal form, and the spatial frame can be rotated to obtain~$\sigma_{ab}={\rm diag}(-2\sigma,\sigma,\sigma)$, $\tau_{ab}={\rm diag}(-2\tau,\tau,\tau)$.
\par 
It is important to note that the Israel-Stewart theory is derived under assumption that the fluid is close to the local equilibrium state. This implies that the relative dissipative fluxes are small:
\begin{equation}
 \vert \pi \vert << p, \qquad (\tau_{ab}\tau^{ab})^{1/2}<<p.
\end{equation}
\par 
Finally, we do not include any mechanisms of particle creation/destruction in the model. The particle current density is then conserved:
\begin{equation}
(nu^\alpha)_{;\alpha} = 0,
\end{equation}

\par 
The dynamics of the dissipative fluxes is described by the Israel-Stewart transport equations, the dots representing derivation with respect to the cosmological time~$t$:
\begin{align}
\tau_0 \dot {\pi} + \pi &= -3\zeta H -\frac{1}{2}\tau_0 \pi \left[ 3H+\frac{\dot{\tau}_0}{\tau_0}-\frac{\dot{\zeta}}{\zeta}-\frac{\dot{T}}{T} \right],\\
\tau_2 \dot{\tau} + \tau &= -2\eta \sigma -\frac{1}{2}\tau_2 \tau \left[ 3H+\frac{\dot{\tau}_2}{\tau_2}-\frac{\dot{\eta}}{\eta}-\frac{\dot{T}}{T} \right],
\end{align}
where the bulk and shear viscosity coefficients $\zeta$ and $\eta$, respectively, are related to the corresponding relaxation times $\tau_0$ and $\tau_2$ by
\begin{equation}
\tau_0 = \zeta \beta_0, \qquad \tau_2=2 \eta \beta_2.
\end{equation}
The non-negative functions $\beta_0$ and~$\beta_2$ are the transient coefficients for scalar and tensor contributions to the entropy density, respectively.
\par 
The relativistic kinetic theory yields~\cite{Schweizer1982}:
\begin{equation}
\begin{array}{lll}
\displaystyle{ \frac{1}{\beta_0} = 4aT^4 \xi^2,} & \qquad & \displaystyle{ \frac{1}{\beta_2}= \frac{8}{15}aT^4,}\\
\displaystyle{ \zeta = \frac{12aT^4}{\bar{\kappa}_0}\xi^2,} & \qquad & \displaystyle{ \eta	= \frac{4aT^4}{15 \bar{\kappa}_2},}
\end{array}
\end{equation}
where the function
\begin{equation}
\xi = \frac{1}{3}-{\left( \frac{\delta p}{\delta T} \right)}_n {\left( \frac{\delta \rho}{\delta T} \right)}^{-1}_n
\end{equation}
represents the deviation of the fluid mixture from the purely radiative behaviour.
\par 
The coefficients~$\bar{\kappa}_0$ and~$\bar{\kappa}_2$ are defined by:
\begin{equation}
\frac{1}{\bar{\kappa}_0}=R \left( \frac{1}{\kappa_0} \right), \qquad  \frac{1}{\bar{\kappa}_2}=R \left( \frac{1}{\kappa_2} \right),
\end{equation}
where $R$ stands for the Rosseland mean of a function, see Appendix. The functions $\kappa_0$ and~$\kappa_2$ can, in principle, be exactly determined by means of the relativistic kinetic theory, provided that a particular model for the material medium is chosen, see~\cite{Schweizer1982} for an example of such a calculation. 
\par 
In the present research we construct a more general model. It follows from the form of the (linearized) collision term of the relativistic Boltzmann equation \cite{Schweizer1982} that~$\kappa_0,\kappa_2 \propto n$. This allows to define
\begin{equation}
\bar{\kappa}_0 = 3s_0mn, \qquad \bar{\kappa}_2 = s_2mn.
\end{equation}
Here the constants~$s_0$ and~$s_2$ are the result of integration~(\ref{Eq:A1:RosselandInt}), provided the integrals converge, and represent some "effective" interaction cross-sections divided by the material particle mass~$m$. The dimension of~$s_0$ and~$s_2$ is that of~$H^{-1}$.

%%%%%%%%%%%%%%%%%%%%%%%%%%%%%%%%%%%%%%%%%%%%%%%%%%%%%%%%%%%%

\section*{THE ENERGY VARIABLES}
We decompose the energy density variable as~$\rho = \epsilon + u + r$, where the terms are defined by:
\begin{equation}
\epsilon = mn, \qquad u=\frac{1}{3}nT, \qquad r=aT^4.
\end{equation}
The Einstein field equations (in terms of scalars) now read:
\begin{align}
\dot{H} &= -H^2 -2\sigma^2-\frac{1}{6}\left(\epsilon+10u+2r+ 3 \pi \right), \\
\dot{\sigma} &= -3H\sigma + \tau, \\
H^2 &= \frac{1}{3}\left( \epsilon +u +r \right) + \sigma^2.
\end{align}
As the next step, we introduce 
\begin{equation}
\theta = \frac{\dot{T}}{T}+H=-\frac{8Hu+3H\pi+6\sigma\tau}{u+4r}
\end{equation}
and calculate
\begin{equation}
\xi = -\frac{8u}{3(u+4r)}.
\end{equation}
Then, the equations for the fluid are:
\begin{align}
\dot{\epsilon} &= -3H\epsilon, \\
\dot{u} &= (-4H+\theta)u, \\
\dot{r} &= 4(-H+\theta) r,\\
\dot {\pi} &= -\left[4H - \frac{5}{2}\theta +\frac{12r\theta}{u+4r}+s_0 \epsilon \right]\pi -12Hr\xi^2,\\
\dot{\tau} &= -\left[4H-\frac{5}{2}\theta + s_2\epsilon \right]\tau -\frac{8}{15}r\sigma.
\end{align}

%%%%%%%%%%%%%%

\section*{THE DIMENSIONLESS FORMULATION}
It is convenient to rewrite the dynamical system in terms of dimensionless, scale-independent variables. The dimensionless time~$\tilde{t}$ is defined by
\begin{equation}
\frac{\rm{d}\tilde{t}}{\rm{d}t}=H, 
\end{equation}
so that $\tilde{t}\to \infty $ as $t\to \infty$ in all ever-expanding cosmological models. Then, Hubble-normalized variables are introduced by:
\begin{align}
(\Sigma, \Theta) &= (\sigma, \theta)/H,\\
(E, U, R, \Pi, \mathcal{T}) &= (\epsilon, u, r, \pi, 3\tau)/3H^2,\\
(S_0, S_2) &= (s_1, s_2)\times H. \label{Eq:Eq:DL-S}
\end{align}
The evolution equations for the cross-section terms~$S_0$ and~$S_2$ follow immediately from (\ref{Eq:Eq:DL-S}). The complete dynamical system can now be written as follows (the primes denote derivation with respect to dimensionless time~$\tilde{t}$):
\begin{align}
\Sigma^\prime &= (q-2)\Sigma + \mathcal{T}, \label{Eq:Eq:DL:Sigma} \\ 
E^\prime &= (2q-1)E,\\
U^\prime &= \left[ 2(q-1)+\Theta \right]U,\\
R^\prime &= 2(q-1+2\Theta)R,\\
\Pi^\prime &= \left[ 2(q-1)+\frac{5}{2}\Theta - \frac{12R\Theta}{U+4R}-S_0E \right]\Pi - 12\xi^2 R,\\
\mathcal{T}^\prime &= \left[ 2(q-1)+\frac{5}{2}\Theta -S_2E \right]\mathcal{T}-\frac{8}{5}R\Sigma,\\
S_0^\prime &= -(q+1)S_0, \\
S_2^\prime &= -(q+1)S_2, \label{Eq:Eq:DL-S2} \\
1 &= E+U+R+\Sigma^2, \label{Eq:Eq:Hamilton}
\end{align}
with 
\begin{align}
q &= 2\Sigma^2 + \frac{1}{2}E+5U+R+\frac{3}{2}\Pi, \label{Eq:Eq:DL-q}\\
\Theta &= -\frac{8U+3\Pi+2\Sigma \mathcal{T}}{U+4R},\\
\xi &= -\frac{8}{3}\cdot \frac{U}{U+4R}, \label{Eq:Eq:DL-chi}
\end{align}
where the so-called deceleration parameter 
\begin{equation}
q \equiv - \dot{H}/H^2-1
\end{equation}
is negative when the cosmological expansion is accelerated, and vice versa.
\par
The Hamiltonian constraint (\ref{Eq:Eq:Hamilton}) allows to exclude~$E$ from the list of variables. The dimension of the physical state space is then seven, the state vector being
\begin{equation}
\mathbf{X}=[\Sigma,U,R,\Pi,\mathcal{T},S_0,S_2].
\end{equation}
The variables $\Sigma,~U$ and~$R$ are bounded, since $U>0$,~$R>0$ and~${\Sigma^2+U+R<1}$; the cross-section variables are subject to $S_0,~S_2\geq 0$; at the same time, no mathematical restriction is imposed on the variables~$\Pi$ and~$\mathcal{T}$.  We also note that although the variables~$U$ and~$R$ are in fact {\it not} independent, it is convenient to treat them as such, considering by this a mathematically more general system. In this case, the state vector belongs to a subspace of~$S^4\times \mathbb{R}^4$.
\par 
We perform complete numerical integrations of the system (\ref{Eq:Eq:DL:Sigma})-(\ref{Eq:Eq:DL-S2}) modulo the Hamiltonian constraint~(\ref{Eq:Eq:Hamilton}) at different sets of initial conditions. In numerical runs, we choose to satisfy~(\ref{Eq:Eq:Hamilton}) initially.

%%%%%%%%%%%%%%

\section*{THE SYSTEM FOR ANALYTICAL INVESTIGATIONS}
Note that the right-hand sides of the equations contain terms which can possibly diverge when~$U,R \to 0$. To resolve this problem, we introduce the new variables~$X,Y,Z$ by
\begin{equation}
U=RX, \qquad \Pi=RY, \qquad \mathcal{T}=RZ.
\end{equation}
The dynamical system is then rewritten as follows:
\begin{align}
\Sigma^\prime &= (q-2)\Sigma + RZ, \label{Eq:Eq:XYZstart}\\
R^\prime &= 2(q-1+2\Theta)R,\\
X^\prime &= -3\Theta X,\\
Y^\prime &= -\left[ \frac{3}{2}\Theta + \frac{12\Theta}{X+4}+S_0E\right]Y-12\xi^2,\\
Z^\prime &= -\left[\frac{3}{2}\Theta +S_2E \right] Z-\frac{8}{5}\Sigma,\\
S_0^\prime &= -(q+1)S_0,\\
S_2^\prime &= -(q+1)S_2,
\end{align}
with 
\begin{align}
q &= 2\Sigma^2+\frac{1}{2}E+5RX+R+\frac{3}{2}RY,\\
E &= 1-RX-R-\Sigma^2,\\
\Theta &= -\frac{8X+3Y+2\Sigma Z}{X+4},\\
\xi &= -\frac{8X}{3(X+4)} \label{Eq:Eq:XYZend}.
\end{align}

Note that~$X>0$ by definition. Therefore, the system~(\ref{Eq:Eq:XYZstart})-(\ref{Eq:Eq:XYZend}) does not contain any singular terms and can be used for analytical investigations, in particular for the local stability analysis.

%%%%%%%%%%%%%%%%%%%%%%%%%%%%%%%%%%%%%%%%%%%%%%%%%%%%%%%%

\section*{RESULTS AND DISCUSSION}

Calculating the eigenvalues corresponding to all the fixed points of the dynamical system reveals that the only stationary point of the system acting as a possible local attractor in the future is
\begin{equation}
\label{Eq:Results:Attractor}
[\Sigma,R,X,Y,Z,S_0,S_2]=0, \qquad {\rm with}~E=1~{\rm and}~q=1/2.
\end{equation}
Extensive numerical runs do not indicate the existense of other kinds of attractors, e.\,g. attracting curves. This allows us to conjecture that (\ref{Eq:Results:Attractor}) is actually the {\it global} attractor of the system. 
\par 
This attractor corresponds to an asymptotically isotropic cosmological model, which is ultimately dominated by the material component. The expansion of the model is decelerated in the future; thus, the transport properties of the physical fluid alone do not provide a mechanism for accelerating the expansion of the Universe.
\par 
The eigenalues corresponding to the state~(\ref{Eq:Results:Attractor}) are~$(-3/2,-1,0,0,0,-3/2,-3/2)$. A careful analysis shows that the attractor is locally stable in the future, the asymptotic decay rates of the variables being:
\begin{align}
\Sigma &\propto \tilde{t}^{5/6}e^{-\tilde{t}},\\
R & \propto \tilde{t}^{4/3}e^{-\tilde{t}},\\
X & \propto \tilde{t}^{-1},\\
Y & \propto \tilde{t}^{-1},\\
Z & \propto \tilde{t}^{-1/2},\\
S_0 & \propto e^{-3/2\tilde{t}},\\
S_2 & \propto e^{-3/2\tilde{t}},
\end{align}
which implies that $[U,\Pi,\mathcal{T}]\to 0$, with
\begin{align}
U & \propto \tilde{t}^{1/3}e^{-\tilde{t}},\\
\Pi & \propto \tilde{t}^{1/3}e^{-\tilde{t}},\\
\mathcal{T} & \propto \tilde{t}^{5/6}e^{-\tilde{t}}.
\end{align}
The constants of proportionality, as well as the order of the correction terms, can be determined by analyzing the centre manifold, which in this case is of dimension three; see~\cite{Coley2008} for an application to a similar dynamical system.
\par 
The relative dissipative fluxes caused by the bulk and shear viscous stresses, respectively, decay in the future:
\begin{align}
\frac{ \vert \pi \vert }{p}  & \propto \tilde{t}^{-1},\\
\frac{(\tau_{ab}\tau^{ab})^{1/2}}{p} & \propto \tilde{t}^{-1/2}.
\end{align}
So, the two-component fluid approaches thermal equilibrium in the asymptotic future, although it happens rather slowly due to the cosmic expansion.

%%%%%%%%%%%%%%%%%%%%%%%%%%%%%%%%%%%%%%%%%%%%%%%%%%%%%%%%%%%%%%%%

\section*{CONCLUSIONS}

We have proposed a mathematical approach which allows to analyze two-component radiative fluids in anisotropic spacetimes using a dynamical systems method, and applied this approach to Bianchi type~I cosmological models. We have used the full Israel-Stewart theory of irreversible thermodynamics to describe the dissipative properties of the fluid, having paid attention to the magnitude of the relative dissipative fluxes. We have determined the future attractor of the system of equations governing the dynamics of the cosmological model and found the future asymptotic behaviour of the geometrical and physical variables. 
\par 
All the solutions obtained describe an izotropizing cosmological model dominated by the material component at late times. The viscous stresses decay in the future, making no contribution to accelerating the expansion of the universe. This is essentially different from the cosmological models with mathematical fluids, where the bulk viscous effects can lead to accelerated expansion, see e.\,g.~\cite{Shogin2015,Shogin2016}.
\par 
We have investigated the dynamics of the relative dissipative fluxes, which describe the deviation of the cosmological fluid from the state of local thermal equilibrium. We have found that these fluxes decay in the future; so, the two-component radiative fluid approaches equilibrium at late times. This means that the underlying assumptions of the Israel-Stewart theory are not violated, and the theory itself is valid for the current application. For the simplified mathematical fluids studied earlier, the situation is opposite; namely, a non-vanishing bulk viscosity has been shown to cause a finite, generally not small, dissipative flux, which in turn leads to a breakdown of the Israel-Stewart theory already in simplest anisotropic backgrounds~\cite{Shogin2016}.
\par 
We have found that the currently observed accelerated expansion of the Universe cannot be explained by the transport phenomena in the radiative fluid alone. However, the spatial Bianchi type~I anisotropy is eliminated due to the dissipative effects in the fluid. As a whole, the dynamics of the considered fluid model is physically more realistic than that of the mathematical fluid models. We therefore conclude that in describing the transport phenomena in early Universe, the physics-based fluid models must be preferred over the simplified mathematical ones.

\bibliography{Bib/New}

\appendix 

\section*{APPENDIX: THE ROSSELAND MEANS}
The Rosseland mean of a frequency-dependent function $\phi(\omega)$ is by definition:
\begin{equation}
\label{Eq:A1:RosselandInt}
R(\phi)=I^{-1}\int \limits_0^\infty {\rm d}\omega \omega^4 F^{(0)}(1+F^{(0)})\phi,
\end{equation}
where 
\begin{equation}
I=\int \limits_0^\infty {\rm d}\omega \omega^4 F^{(0)}(1+F^{(0)})=4\pi^2aT^5,
\end{equation}
and $F^{(0)}$ is the equilibrium distribution function for the radiation quanta. In particular, for photons it is given by
\begin{equation}
F^{(0)}(\omega)={\left( e^{-\omega/T}-1 \right)}^{-1}.
\end{equation}

\end{document}